\documentclass{article}
\usepackage{spconf,amsmath,graphicx}

\usepackage{times}
\usepackage[latin1]{inputenc}
\usepackage[T1]{fontenc}
\usepackage[german,greek,english]{babel}
\usepackage{epsfig}
\usepackage{amsbsy}
\usepackage{amsmath}
\usepackage{fixmath}
\usepackage{graphicx}
\usepackage{subfigure}
\usepackage{amssymb}
\usepackage{latexsym}
\usepackage[dvipsnames,usenames]{color}
\usepackage{framed, color}
\usepackage{verbatim}
\usepackage{algorithm}
\usepackage{algorithmic}

\usepackage{tikz}
\usepackage{pgfplots}
\pgfplotsset{compat=1.3}
\usetikzlibrary{arrows,snakes,shapes,positioning,arrows,decorations.markings,calc}

\tikzset{box_2/.style={rectangle,inner sep=0.5pt,fill=white, minimum 
height=0.1cm,minimum width=0.1cm,draw=white,thin}}

\DeclareMathAlphabet{\mathbit}{OML}{cmr}{bx}{it}
\DeclareMathAlphabet{\mathsf}{OT1}{cmss}{m}{n}
\DeclareMathAlphabet{\mathbsf}{OT1}{cmss}{bx}{it}

\DeclareMathOperator{\E}{E}
\DeclareMathOperator{\T}{T}

\newcommand{\ve}[1]{\mbox{$\mathbit#1$}}

\newcommand{\exin}[2]{\E_{#1} \left[#2\right]}

\title{Asymptotic Performance Analysis for 1-bit Bayesian Smoothing}
\name{Lin Zhang, Manuel Stein, Josef A. Nossek}
\address{Institute for Circuit Theory and Signal Processing, Technische Universit\"at M\"unchen, Germany\\
E-mail: lin.zhang@tum.de, manuel.stein@tum.de, josef.a.nossek@tum.de}
\begin{document}
%
\maketitle
\begin{abstract}
Energy-efficient signal processing systems require estimation methods operating on data collected with low-complexity devices. Using analog-to-digital converters (ADC) with $1$-bit amplitude resolution has been identified as a possible option in order to obtain low power consumption. The $1$-bit performance loss, in comparison to an ideal receiver with $\infty$-bit ADC, is well-established and moderate for low SNR applications ($2/\pi$ or $-1.96$ dB). Recently it has been shown that for parameter estimation with state-space models the $1$-bit performance loss with Bayesian filtering can be significantly smaller ($\sqrt{2/\pi}$ or $-0.98$ dB). Here we extend the analysis to Bayesian smoothing where additional measurements are used to reconstruct the current state of the system parameter. Our results show that a $1$-bit receiver performing smoothing is able to outperform an ideal $\infty$-bit system carrying out filtering by the cost of an additional processing delay $\Delta$.
\end{abstract}
\section{Introduction}\label{sec:intro}
Parameter estimation finds its application in diverse problems like radar, communications and robotics, where one faces the problem of measuring the current state of the system parameters from measurements. Due to the fact that mobile processing systems are required to operate under strict power and hardware constraints, using measurement data that is obtained with coarse resolution ADCs has gained attention \cite{Lok98,Madsen00,Mez10}. In particular, a 1-bit ADC provides a sampling device which can be operated energy-efficient at high sampling rates \cite{Walden99}. The non-linear operation of $1$-bit hard-limiting results in a performance loss which is moderate in the low SNR regime ($2/\pi$ or $-1.96$ dB) \cite{Vleck66}. As in many applications the parameters change slowly compared to the sampling rate, they can be assumed to be constant within one measurement block and to vary slightly from block to block. If the evolution of the parameters is taken into account by formulating a state-space model and performing filtering \cite{Clements72,Sviestins00,Karlsson05,Duan2008_2,Duan2008}, it turns out that the performance loss can be reduced to $-0.98$ dB when using a 1-bit quantizer in the low SNR regime \cite{Stein15}. 

Pursuing this line of work, we extend the $1$-bit performance analysis to Bayesian smoothing. Besides the observations of passed blocks, with smoothing $\Delta$ additional measurements are taken into account and the estimate is therefore provided with a delay of at least $\Delta$. In practice technical requirements specify a delay budget which can be used in order to perform all required calculations. The simplicity of $1$-bit ADCs and the fact that the resulting binary receive signals allow to speed up the digital hardware, enables to use a portion of the delay budget to acquire $\Delta$ additional measurements. In order to reduce the mean square error (MSE), these observations can be incorporated into the estimation algorithm by implementing smoothing.

Our discussion on the performance of $1$-bit smoothing is based on Bayesian bounds for the achievable MSE. After the definition of a generic system model, we outline the basic concepts of Bayesian filtering, prediction and smoothing. A recursive Bayesian Cram\'{e}r-Rao lower bound (BCRLB) is reviewed for these methods. Then we compare the performance of the $1$-bit system with an ideal $\infty$-bit receiver. Finally, we show the performance gain which is possible with $1$-bit smoothing in comparison to $\infty$-bit filtering.
\section{Problem Statement}
\subsection{System Model}
For the discussion we assume $k = 1, 2, \ldots, K$ receive blocks with the receive signal $\mbox{\boldmath$y$}_k\in\mathbb{R}^N$ in the $k$-th block following the measurement probability density function (PDF) $p(\mbox{\boldmath$y$}_k|\mbox{\boldmath$\theta$}_{k})$, where $\mbox{\boldmath$\theta$}_{k}\in\mathbb{R}^M$ is the $k$-th block parameter. The parameter follows a Markov model \cite{Saerkkae13}, described by a transition PDF $p(\mbox{\boldmath$\theta$}_k|\mbox{\boldmath$\theta$}_{k-1})$. We assign a prior PDF $p(\mbox{\boldmath$\theta$}_0)$ to the parameter vector before the inital measurement block $k=1$. 
\subsection{Bayesian Filtering}
The goal of filtering is to calculate the filtering PDF $p\left(\ve{\theta}_k|\ve{Y}_k\right)$ of the current state $\mbox{\boldmath$\theta$}_k$ given the measurement matrix 
\begin{align}
\ve{Y}_k=\left[\ve{y}_1 \quad \ve{y}_2 \quad \ldots \quad \ve{y}_k\right] \in\mathbb{R}^{N\times k},
\end{align}
containing all of the received samples up to block $k$. From the filtering PDF the conditional mean estimator (CME) 
\begin{align}
\hat{\ve{\theta}}_{k|k}(\ve{Y}_k)=\exin{\ve{\theta}_k|\ve{Y}_k}{\ve{\theta}_k}
\end{align}
can be deduced. The CME minimizes the filtering MSE
\begin{align}
\ve{B}_{k|k}=\exin{\ve{\theta}_k,\ve{Y}_{k}}{\big(\ve{\theta}_k-\hat{\ve{\theta}}_{k|k}(\ve{Y}_k)\big)\big(\ve{\theta}_k-\hat{\ve{\theta}}_{k|k}(\ve{Y}_k)\big)^{\T}}.
\end{align}
\subsection{Bayesian Prediction}
Bayesian prediction is referred to as the inference of the parameter $\hat{\ve{\theta}}_{l|k}$ in the $l$-th block with $k<l$ \cite{Chen03}. The dynamic model of the parameter $p(\mbox{\boldmath$\theta$}_k|\mbox{\boldmath$\theta$}_{k-1})$ is decisive under these circumstances. The CME is calculated from $p\left(\ve{\theta}_l|\ve{Y}_k\right)$ by
\begin{align}\label{prediction:estimator}
\hat{\ve{\theta}}_{l|k}(\ve{Y}_k)=\exin{\ve{\theta}_l|\ve{Y}_k}{\ve{\theta}_l}
\end{align}
and minimizes the MSE
\begin{align}\label{prediction:mse}
\ve{B}_{l|k}=\exin{\ve{\theta}_l,\ve{Y}_{k}}{\big(\ve{\theta}_l-\hat{\ve{\theta}}_{l|k}(\ve{Y}_k)\big)\big(\ve{\theta}_l-\hat{\ve{\theta}}_{l|k}(\ve{Y}_k)\big)^{\T}}.
\end{align}
\subsection{Bayesian Smoothing}
Bayesian smoothing reconstructs the state $\hat{\mbox{\boldmath$\theta$}}_{l|k}$ in block $l$ by taking into account future measurements $\ve{Y}_k$ up to the $k$-th block \cite{Chen03}. The expression for the CME and the MSE is the same as in \eqref{prediction:estimator} and \eqref{prediction:mse} while for smoothing $l<k$.
\section{Bayesian Cram\'{e}r-Rao Bounds}\label{sec:pagestyle}
\subsection{Filtering Bayesian Cram\'{e}r-Rao Lower Bound}
For Bayesian filtering the MSE can be bounded by the Bayesian Cram\'{e}r-Rao lower bound \cite{Ristic04,Trees07}
\begin{align}
\ve{B}_{k|k}\succeq\ve{J}_{k|k}^{-1}.
\end{align}
The required filtering Bayesian information matrix (BIM) $\ve{J}_{k|k}$ can be calculated efficiently in a recursive manner \cite{Tichavsky98} 
\begin{align}\label{recursion:filtering}
\ve{J}_{k|k}=\ve{D}_{k}^{22}+\exin{\ve{\theta}_{k}}{\ve{F}(\ve{\theta_{k}})}-\ve{D}_{k}^{21}(\ve{J}_{k-1|k-1}+\ve{D}_{k}^{11})^{-1}\ve{D}_{k}^{12}
\end{align}
with
\begin{align}
\ve{D}_{k}^{11}&=\exin{\ve{\theta}_{k-1},\ve{\theta}_k}{\frac{\partial \operatorname{ln}p(\ve{\theta}_{k}|\ve{\theta}_{k-1})}{\partial \ve{\theta}_{k-1}}\bigg(\frac{\partial \operatorname{ln}p(\ve{\theta}_{k}|\ve{\theta}_{k-1})}{\partial \ve{\theta}_{k-1}}\bigg)^{\T}}\\
\nonumber\ve{D}_{k}^{12}&=\exin{\ve{\theta}_{k-1},\ve{\theta}_k}{\frac{\partial \operatorname{ln}p(\ve{\theta}_{k}|\ve{\theta}_{k-1})}{\partial \ve{\theta}_{k-1}}\left(\frac{\partial \operatorname{ln} p(\ve{\theta}_{k}|\ve{\theta}_{k-1})}{\partial \ve{\theta}_{k}}\right)^{\T}}\\
&={\ve{D}_{k}^{21}}^\top\\
\ve{D}_{k}^{22}&=\exin{\ve{\theta}_{k-1},\ve{\theta}_k}{\frac{\partial \operatorname{ln}p(\ve{\theta}_{k}|\ve{\theta}_{k-1})}{\partial \ve{\theta}_{k}}\bigg(\frac{\partial \operatorname{ln}p(\ve{\theta}_{k}|\ve{\theta}_{k-1})}{\partial \ve{\theta}_{k}}\bigg)^{\T}}
\end{align}
and the Fisher information matrix having the entries \cite{Kay93}
\begin{align}
\left[\ve{F}(\ve{\theta}_k)\right]_{ij}=\exin{\ve{y}_k|\ve{\theta}_k}{\frac{\partial\operatorname{ln}p(\ve{y}_k|\ve{\theta}_k)}{\partial\theta_i}\frac{\partial\operatorname{ln}p(\ve{y}_k|\ve{\theta}_k)}{\partial\theta_j}}.
\end{align}
The initial filtering BIM can be computed with the prior PDF
\begin{align}
\ve{J}_{0|0}=\exin{\ve{\theta}_0}{\bigg(\frac{\partial \operatorname{ln}p(\ve{\theta}_{0})}{\partial \ve{\theta}_{0}}\bigg)^2}.
\end{align}
\subsection{Prediction Bayesian Cram\'{e}r-Rao Lower Bound}
The performance of prediction can be lower bounded by
\begin{align}\label{bcrlb:prediction}
\ve{B}_{l|k}\succeq\ve{J}_{l|k}^{-1}, \quad\quad l>k.
\end{align}
The recursive calculation of the prediction BIM is obtained with a similar routine like the filtering BIM \cite{Simandl01}, i.e.,
\begin{align}\label{recursion:prediction}
\ve{J}_{l|k}=\ve{D}_l^{22}-\ve{D}_l^{12}(\ve{D}_l^{11}+\ve{J}_{l-1|k})^{-1}{\ve{D}_l^{21}}.
\end{align}
The recursion \eqref{recursion:prediction} starts from the filtering BIM $\ve{J}_{k|k}$.
\subsection{Smoothing Bayesian Cram\'{e}r-Rao Lower Bound}
The smoothing MSE can be lower bounded by \cite{Simandl01}
\begin{align}
\ve{B}_{l|k}\succeq\ve{J}_{l|k}^{-1}, \quad\quad l<k,
\end{align}
where $\ve{J}_{l|k}$ is the smoothing BIM with backward recursion 
\begin{align}\label{recursion:smoothing}
\ve{J}_{l|k}&=\ve{J}_{l|l}+\ve{D}_{l+1}^{11}\notag\\
&-\ve{D}_{l+1}^{21}(\ve{D}_{l+1}^{22}+\ve{J}_{l+1|k}-\ve{J}_{l+1|l})^{-1}{\ve{D}_{l+1}^{12}}
\end{align}
which starts with the filtering BIM $\ve{J}_{k|k}$. Substituting the one-step version $\ve{J}_{k+1|k}$ of the prediction BIM \eqref{recursion:prediction} into  \eqref{recursion:smoothing} a compact form of the smoothing BIM can be obtained
\begin{align}
\ve{J}_{l|k}=\ve{J}_{l|l}+\ve{\kappa}(l|k),
\end{align}
where the smoothing gain $\ve{\kappa}(l|k)$ follows the recursive rule
\begin{align}
\nonumber\ve{\kappa}(l|k)&=\ve{D}_{l+1}^{11}-\ve{D}_{l+1}^{21}(\ve{D}_{l+1}^{22}
\\&+\exin{\ve{\theta}_{l+1}}{\ve{F}(\ve{\theta}_{l+1})}+\ve{\kappa}(l+1|k))^{-1}{\ve{D}_{l+1}^{12}}
\end{align}
with the initial value
\begin{align}
\ve{\kappa}(k-1|k)=\ve{D}_{k}^{11}-\ve{D}_{k}^{21}(\ve{D}_{k}^{22}+\exin{\ve{\theta}_{k}}{\ve{F}(\ve{\theta}_{k})})^{-1}{\ve{D}_{k}^{12}}.
\end{align}
Note that the smoothing BIM $\ve{J}_{l|k}$ is the filtering BIM $\ve{J}_{l|l}$ plus the gain factor $\ve{\kappa}(l|k)$, which represents the additional information obtained by using smoothing instead of filtering.
\section{Gaussian Random Walk}
For the performance analysis we consider a Gaussian random walk model. For the $1$-bit receiver the information about the amplitude of the measurement is discarded.
\subsection{Gaussian Random Walk}
The unquantized measurement model of the random walk is 
\begin{align}
y_k=\theta_k + \eta_k,
\end{align}
where we assume that $\eta_k\sim\mathcal{N}(0,\sigma_{\eta}^2)$. The parameter $\theta_k$ evolves over time by following the autoregressive model
\begin{align}\label{rand:walk:state:space}
\theta_k=\alpha\theta_{k-1}+z_k,
\end{align}
where $z_k\sim\mathcal{N}(0,\sigma_z^2)$. The prior distribution $p(\theta_0)$ is assumed to be Gaussian, i.e. $\theta_{0}\sim\mathcal{N}(\mu_0,\sigma_0^2)$. The required expressions for \eqref{recursion:filtering}, \eqref{recursion:prediction} and \eqref{recursion:smoothing} are
\begin{align}
D_k^{11}=\frac{\alpha^2}{\sigma_z^2},\quad\quad D_k^{22}=\frac{1}{\sigma_z^2},\quad\quad D_k^{12}=D_k^{21}=-\frac{\alpha}{\sigma_z^2}.
\end{align}
and
\begin{align}
\exin{\theta_k}{F(\theta_k)}=\frac{1}{\sigma_\eta^2}.
\end{align}
The filtering BIM can be explicitly written as 
\begin{align}
J_{k|k}=\frac{1}{\sigma_z^2}+\frac{1}{\sigma_\eta^2}-\left(\frac{\alpha}{\sigma_z^2}\right)^2\left(J_{k-1|k-1}+\frac{\alpha^2}{\sigma_z^2}\right)^{-1}.
\end{align}
For prediction ($l>k$) the BIM is
\begin{align}
J_{l|k}=\frac{1}{\sigma_z^2}-\left(\frac{\alpha}{\sigma_z^2}\right)^2\left(\frac{\alpha^2}{\sigma_z^2}+J_{l-1|k}\right)^{-1},
\end{align}
while for smoothing ($l<k$) the BIM is given by
\begin{align}
J_{l|k}=J_{l|l}+\kappa(l|k)
\end{align}
with the smoothing gain
\begin{align}
\kappa(l|k)=\frac{\alpha^2}{\sigma_z^2}-\left(\frac{\alpha}{\sigma_z^2}\right)^2\left(\frac{1}{\sigma_z^2}+\frac{1}{\sigma_\eta^2}+\kappa(l+1|k)\right)^{-1}.
\end{align}
\subsection{$1$-bit Gaussian Random Walk}
The $1$-bit receiver operates on a hard-limited measurement
\begin{align}
r_k=\operatorname{sign}(\theta_k+\eta_k).
\end{align}
Under this circumstances the expected FIM
\begin{align}\label{efim:quantized}
\exin{\theta_k}{F_q(\theta_k)}=\int\limits_{-\infty}^{\infty} F_q(\theta_k)p_{\theta_k}(\theta_k)\mathrm{d}\theta_k
\end{align}
has to be calculated with
\begin{align}
F_q(\theta_k)=\frac{1}{2\pi\sigma_\eta^2}\frac{\operatorname{e}^{-\frac{\theta_k^2}{\sigma_\eta^2}}}{\operatorname{Q}\big(\frac{\theta_k}{\sigma_\eta}\big)\operatorname{Q}\big(-\frac{\theta_k}{\sigma_\eta}\big)},
\end{align}
where $\operatorname{Q}(\cdot)$ is the Q-function. For the evaluation of \eqref{efim:quantized}, note that the mean and variance of $\theta_k$ evolve according to \cite{Stein15}
\begin{align}
\exin{\theta_k}{\theta_k}&=\alpha^k\mu_0\\
\operatorname{Var}\left[\theta_k\right]&=\alpha^{2k}\sigma_0^2+\left(\sum\limits_{i=1}^{i=k}\alpha^{2(k-i)}\right)\sigma_z^2,
\end{align}
such that
\begin{align}
\sigma_{\theta,\infty}^{2} = \lim_{k\rightarrow\infty} \operatorname{Var}\left[\theta_k\right] = \frac{1}{1-\alpha^2} \sigma_z^2.
\end{align}
The filtering BIM for the quantized random walk is
\begin{align}
J_{k|k,q}=\frac{1}{\sigma_z^2}+\exin{\theta_k}{F_q(\theta_k)}-\left(\frac{\alpha}{\sigma_z^2}\right)^2\left(J_{k-1|k-1,q}+\frac{\alpha^2}{\sigma_z^2}\right)^{-1}.
\end{align}
For prediction ($l>k$) the quantized BIM is
\begin{align}
J_{l|k,q}=\frac{1}{\sigma_z^2}-\left(\frac{\alpha}{\sigma_z^2}\right)^2\left(\frac{\alpha^2}{\sigma_z^2}+J_{l-1|k,q}\right)^{-1},
\end{align}
whereas for smoothing ($l<k$) the quantized BIM is
\begin{align}
\quad J_{l|k,q}&=J_{l|l,q}+\kappa_{q}(l|k)
\end{align}
with
\begin{align}
\kappa_{q}(l|k)&=\frac{\alpha^2}{\sigma_z^2}-\left(\frac{\alpha}{\sigma_z^2}\right)^2\cdot
\\&\left(\frac{1}{\sigma_z^2}+\exin{\theta_{l+1}}{F_q(\theta_{l+1})}-J_{l+1|l+1,q}+J_{l+1|k,q}\right)^{-1}.
\end{align}
\subsection{Unquantized vs. Quantized Random Walk}
For the comparison of the performance obtained with the unquantized and the quantized random walk model, we consider the asymptotic performance of the system. After an initial transient phase the filtering algorithm reaches a steady-state, where the estimation error saturates \cite{Stein15}
\begin{align}
&J_{k|k,\infty}=J_{k+1|k+1,\infty},\\
&J_{k|k,q}=J_{k+1|k+1,q}.
\end{align}
Therefore, we define the asymptotic filtering BIMs as follows
\begin{align}
\tilde{J}_{k|k,\infty}&=\lim_{k\rightarrow\infty} J_{k|k,\infty},\\
\tilde{J}_{k|k,q}&=\lim_{k\rightarrow\infty} J_{k|k,q}.
\end{align}
Also for Bayesian smoothing a steady-state is obtained for the unquantized and the quantized system , where
\begin{align}
&J_{k|k+\Delta,\infty}=J_{k+1|k+1+\Delta,\infty},\\
&J_{k|k+\Delta,q}=J_{k+1|k+1+\Delta,q}.
\end{align}
Therefore, we define the asymptotic smoothing BIMs by
\begin{align}
\tilde{J}_{k|k+\Delta,\infty}=&\lim_{k\rightarrow\infty,\Delta\rightarrow\infty} J_{k|k+\Delta,\infty},\\
\tilde{J}_{k|k+\Delta,q}=&\lim_{k\rightarrow\infty,\Delta\rightarrow\infty} J_{k|k+\Delta,q}.
\end{align}
The $1$-bit quantization loss of Bayesian smoothing in steady-state can be defined by the information ratio
\begin{align}\label{info:ratio:smoothing}
\rho_{\text{SL}}=\frac{\tilde{J}_{k|k+\Delta,q}}{\tilde{J}_{k|k+\Delta,\infty}}.
\end{align}
Analogously the $1$-bit quantization loss of Bayesian filtering is denoted by
\begin{align}\label{info:ratio:filtering}
\rho_{\text{F}}=\frac{\tilde{J}_{k|k,q}}{\tilde{J}_{k|k,\infty}}.
\end{align}
For the comparison of the steady-state estimation performance of $1$-bit smoothing with respect to unquantized filtering we introduce the information ratio  
\begin{align}\label{info:gain:smoothing}
\rho_{\text{S}}=\frac{\tilde{J}_{k|k+\Delta,q}}{\tilde{J}_{k|k,\infty}}.
\end{align}

\section{Result}
\label{sec:majhead}
For the visualization of the expression \eqref{info:ratio:smoothing} we choose $\alpha=1-10^{-5}$, $\sigma_{\theta_0}=1$ and $\sigma_{\eta}=1$. The result depicted in Fig.1 shows that the quantization loss applying Bayesian smoothing approaches an asymptotic value of $-0.98\text{ dB}$ in the low SNR domain. The quantization loss caused by using 1-bit smoothing instead of unquantized smoothing is the same as for the application of filtering \cite{Stein15}. The reduction of the performance loss from $-1.96\text{ dB}$ without a state-space model \cite{Vleck66} to $-0.98\text{ dB}$ is due to the additional dynamic model \eqref{rand:walk:state:space}, which is not affected by the $1$-bit quantizer. 
\begin{figure}[!ht]
\centering
\begin{tikzpicture}
  	\begin{axis}[ylabel=$10\log(\rho_{\text{SL}})$,
  			xlabel=$10\log\left(\frac{\sigma_{\theta,\infty}^2}{\sigma_\eta^2}\right)$,
			grid,
			xmin=-40,
			xmax=10,
			ymin=-3,
			ymax=-0.5]
			
    	\addplot[smooth, black] table[x index=0, y index=1] {smoothing_inside.txt};
\draw [densely dashed, line width=0.75pt] (axis cs:-20,-0.98) --(axis 
cs:10,-0.98);
\node[box_2] at(axis cs:-5,-0.98) {\tiny{ -0.98\,dB }};

	\end{axis}
\end{tikzpicture}
\caption{Performance ratio $\rho_{\text{SL}}$ vs. SNR}
\label{fig:bayesian smoothing inside} 
\end{figure}
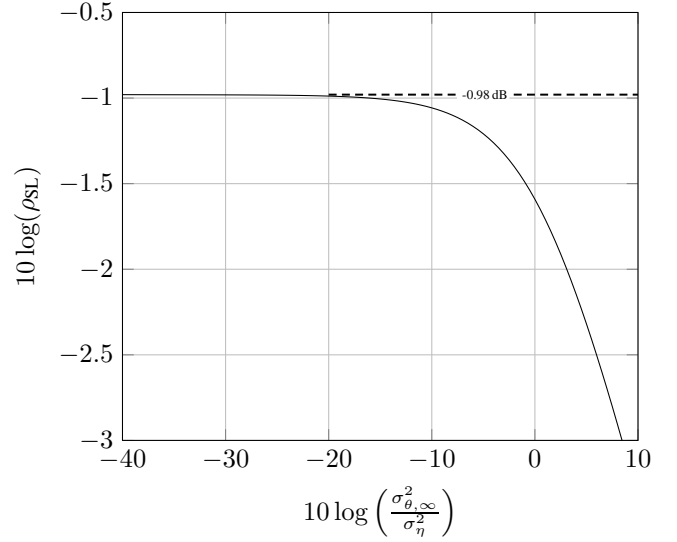
Fig. 2 illustrates the behaviour of the performance ratio \eqref{info:ratio:filtering} and \eqref{info:gain:smoothing} when $\alpha$ approaches one. We can see that $\rho_\text{S}$ is in general larger than $\rho_\text{F}$. The performance ratio $\rho_{\text{S}}$ becomes positive in the low to medium SNR regime for $\alpha$ close to one. This means that the performance loss caused by quantization of the measurements can be compensated by performing smoothing. For the low SNR regime it is possible to outperform the ideal $\infty$-bit filtering algorithm by $2 \operatorname{dB}$ by using a $1$-bit smoothing receiver. However note that this result comes with an additional delay $\Delta$ as more measurements have to be taken into account.
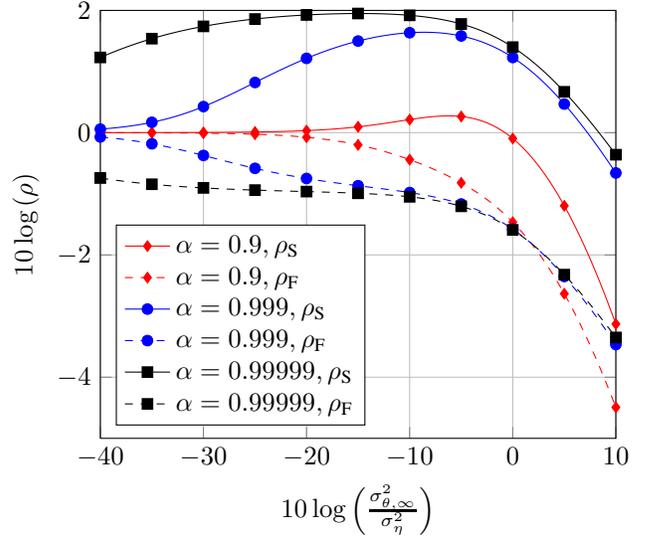
\begin{figure}[!ht]
\centering
\begin{tikzpicture}
  	\begin{axis}[ylabel=$10\log\left(\rho\right)$,
  			xlabel=$10\log\left(\frac{\sigma_{\theta,\infty}^2}{\sigma_\eta^2}\right)$,
			grid,
			xmin=-40,
			xmax=10,
			ymin=-5,
			ymax=2,
			legend cell align=left,
			legend style={legend pos=south west}]
			
		\addplot[smooth, red, every mark/.append style={solid}, mark=diamond*, mark repeat=5] table[x index=0, y index=1] {smoothing0.9.txt};
    	\addlegendentry{$\alpha=0.9,\rho_{\text{S}}$}	
    	
    	\addplot[smooth,red, dashed, every mark/.append style={solid}, mark=diamond*, mark repeat=5] table[x index=0, y index=1] {filtering0.9.txt};
    	\addlegendentry{$\alpha=0.9,\rho_{\text{F}}$}

    	\addplot[smooth,blue, every mark/.append style={solid}, mark=otimes*, mark repeat=5] table[x index=0, y index=1] {smoothing0.999.txt};
    	\addlegendentry{$\alpha=0.999,\rho_{\text{S}}$}
    	
    	\addplot[smooth,blue,dashed, every mark/.append style={solid}, mark=otimes*, mark repeat=5] table[x index=0, y index=1] {filtering0.999.txt};
    	\addlegendentry{$\alpha=0.999,\rho_{\text{F}}$}
    	
    	\addplot[smooth,black, every mark/.append style={solid}, mark=square*, mark repeat=5] table[x index=0, y index=1] {smoothing0.99999.txt};
    	\addlegendentry{$\alpha=0.99999,\rho_{\text{S}}$}
    	
    	\addplot[smooth,black,dashed, every mark/.append style={solid}, mark=square*, mark repeat=5] table[x index=0, y index=1] {filtering0.99999.txt};
    	\addlegendentry{$\alpha=0.99999,\rho_{\text{F}}$}
    	
	\end{axis}
\end{tikzpicture}
\caption{Performance ratio $\rho_{\text{S}}$ and $\rho_{\text{F}}$ vs. SNR ($\alpha \rightarrow 1$)}
\label{fig:gain ratio for alpha tends to 1}
\end{figure}
\section{Conclusion}
\label{sec:print}
We have characterized the effect of coarse quantization onto the state-space estimation performance with smoothing. For a compact analysis the BCRLB has been used in order to approximate the MSEs with efficient processing algorithms. For the smoothing BIM a compact formulation which explicitly contains the smoothing gain factor $\kappa$ has been derived. For the example of a random walk model we carried out a performance analysis for 1-bit Bayesian smoothing by comparing to an ideal system with high measurement resolution. With respect to the performance of $\infty$-bit Bayesian filtering we have found that the $1$-bit quantization-loss can be completely compensated by considering more measurements while performing smoothing. With the example of a random walk, we have shown that it is possible to significantly outperform an unquantized filtering system by a $1$-bit smoothing receiver at the cost of an additional processing delay $\Delta$.
%
%

\end{document}